# Elastically Relaxed Free-standing Strained-Si Nanomembranes

[1]Michelle M. Roberts, [1]Levente J. Klein, [1]Don E. Savage, [1]Keith A. Slinker, [1]Mark Friesen, [2]George Celler, [1]Mark A. Eriksson, [1]Max G. Lagally

[1]University of Wisconsin-Madison, Madison, Wisconsin 53711, USA
[2]Soitec USA, 2 Centennial Dr, Peabody, MA 01960

Strain plays a critical role in the properties of materials. In silicon and silicon-germanium, strain provides a mechanism for control of both carrier mobility and band offsets. In materials integration, strain is typically tuned through the use of dislocations and elemental composition. We demonstrate a versatile method to control strain, by fabricating membranes in which the final strain state is controlled by elastic strain sharing, i.e., without the formation of defects. We grow Si/SiGe layers on a substrate from which they can be released, forming nanomembranes. X-ray diffraction measurements confirm a final strain predicted by elasticity theory. The effectiveness of elastic strain to alter electronic properties is demonstrated by low-temperature longitudinal-Hall effect measurements on a strained-Si quantum well before and after release. Elastic strain sharing and film transfer offers an intriguing path towards complex, multiple-layer structures in which each layer's properties are controlled elastically, without the introduction of undesirable defects.

Strain plays a critical role in the properties of many materials. It is important in devices ranging from high-speed electronics, to micro and nanoelectromechanical systems (MEMS and NEMS), to nanophotonics. In silicon/silicon-germanium semiconductor heterostructures, strain provides control over band offsets and can increase carrier mobility in both Si and SiGe.[1] It is generally desirable to have a Si film with significant tensile strain. Thus, minimizing strain-induced defects, while simultaneously controlling strain, is crucial for obtaining high-quality materials suitable, e.g., for high-performance devices.

The conventional means for producing strain in Si films is to grow the Si on a strain-relaxed SiGe virtual substrate.[2] This virtual substrate, in turn, is created by growing strain-graded SiGe layers on bulk Si, a process that necessarily uses dislocations to provide relaxation.[3] In this process, misfit dislocations at the interfaces between the layers (created by the strain differential between the layers) relax the strain, but, inevitably, they also produce detrimental effects. Every misfit dislocation has two threading arms associated with it that extend through the entire structure, including the ultimate strained-Si layer. Scattering of charge carriers by these threads can degrade the carrier mobility at high thread densities of $10^7$-$10^9$.[4] While lower thread densities (~$10^5$) are obtainable by thick strain grading layers, further reduction may still offer improvement, and complete elimination of the thick strain grading layer in its entirety would be ideal. Misfit dislocations at the strained-Si layer can also extend if this layer exceeds its critical thickness, limiting design freedom.[5,6] Finally, the relaxation process in the SiGe layers generates steps at the sample surface through the opening of dislocation loops. These steps bunch, creating a rough surface referred to as cross-hatch. A polishing and cleaning step is then necessary before further device processing.

Compliant substrates have been proposed as a way to overcome these limitations of dislocation-based relaxation.[7] A compliant substrate, composed of a very thin strained-layer system supported by a thick rigid substrate, is intended to achieve strain control with no dislocations by distributing the strain among the different layers based on their compositions and thicknesses. Compliancy requires, however, that the whole film system can slide freely on the substrate (hence the name compliant), something that has been very difficult to achieve.[8,9] Small-area compliant substrates have used SiGe mesas on borophosphosilicate glass to create relaxed SiGe mesas up to 200 μm x 200 μm.[10]

Extending the concept of compliancy, if the substrate on which the strained-layer system is grown were "nothing," or if the substrate could be removed before the film system is required to relax, one could have all the benefits of a strained-Si film without any of the problems of defect generation or lack of compliancy. This concept has been partially realized with undercut

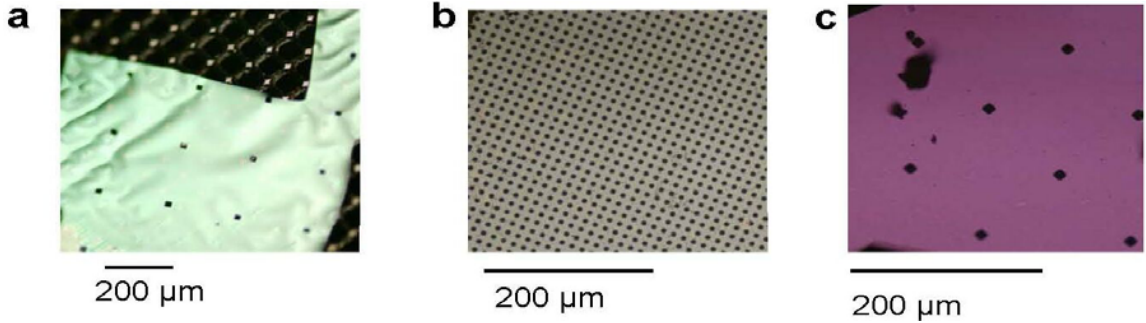

**Figure 1** Silicon nanomembranes transferred onto several substrates. **a**, A metal grid with an L-shaped membrane (light color), showing the flexibility of the membrane. **b**, Glass. **c**, Teflon®.

structures, leaving a substrate supported at discrete points, such as mesas supported either by a center pedestal or at the edges. Undercut substrates have been used for both Si/SiGe and III/V strain-relaxed layer systems.[11-14] However, in this approach relaxation cannot occur effectively at the contact points, and the areas that can be made free-standing are relatively small. Elastically strain relaxed InGaAs regions have been made starting with mesas that were subsequently undercut, allowing the relaxed InGaAs tethered regions to fall to the substrate below to create areas of relaxed material.[15, 16]

We describe here a methodology that extends the above approaches in that it allows the fabrication of large freestanding elastically strain-relaxed membranes. In our case the membranes are SiGe based, but the concept is generalizable. Our membranes are temporarily completely free-floating, and during this step can be transferred to a new substrate. *We can thus effectively create thin strained-Si layers on anything that is not rapidly soluble in water*. We further show that elastic strain can be used to tune the electronic properties of materials. Comparison with theory confirms that the electronic improvements are directly connected to the observed changes in mechanical strain.

We grow SiGe films on ultra-thin Si-on-insulator (SOI) substrates below the thicknesses at which dislocations form. We completely release large continuous membranes of the film system by removing the oxide layer using an HF etch, allowing the film system to achieve the final strain state that minimizes the system free energy. In this paper we focus on membranes transferred to oxidized Si. However, Figures 1a-c show that we can also release membranes as large as 5 mm x 5 mm on metal grids, glass, and Teflon®. Because the final strain state is reached in solution, the final substrate does not play a role in strain management. Nonetheless, strong attachment via van der Waals bonding occurs on all substrates; if stronger chemical attachment is needed for further processing, it must be developed for each new substrate.

It is useful to note that transfer of thin films is not unprecedented – although previous work has not used the transfer process to induce a change in strain. Transfer of thin-film semiconductors to alternate substrates has been achieved using transfer materials such as wax,[17-19] wet printing of small pieces released into water solution,[18] or through wafer bonding of either relaxed or strained films.[20, 21] We believe that these methods are fully compatible with the elastic relaxation methods we present here.

Membrane formation begins using commercially available SOI with typically a 100 nm thick Si template layer and a 200 nm thick buried-oxide layer, but these thicknesses can be larger or smaller. We thin the Si template layer to 20 nm (or any value desired), using a dry thermal oxidation at 1050 ºC, followed by an HF dip. Next, an epitaxial structure is grown consisting of Si, SiGe, and Si layers on the thinned SOI by chemical vapor deposition (CVD) at 580 ºC in a cold-walled ultra-high-vacuum CVD (UHV-CVD) reactor at a pressure of 30 mTorr. We monitor the growth in-situ with reflection high-energy electron diffraction (RHEED).

The composition of the SiGe layers and the thicknesses of all layers can be modified depending on the desired properties. To create flat membranes, we grow films that will balance the strain once the membranes are released. Here we exam two different sets of samples to confirm the presence and effectiveness of elastic-strain sharing. In the first, we grow rather thick films that can be characterized readily with x-ray diffraction, consisting of 48 nm of Si on 128 nm of $Si_{0.84}Ge_{0.16}$ on 56 nm of Si on insulator (SOI). This approximately balanced film structure results, once it is released, in a flat membrane with tensilely strained Si layers sandwiching a compressively strained SiGe layer. In the second set, we deposit a modulation-doped Si quantum well structure sandwiched between $Si_{0.68}Ge_{0.32}$ layers deposited onto a relaxed $Si_{0.79}Ge_{0.21}$-on-insulator (SGOI) substrate. This sample is suited for low-temperature electrical characterization, and also it demonstrates that following growth, yet before release, the samples can be patterned and processed as usual for Si devices. Processing after transfer is also possible, as discussed below. Details of the membrane transfer technique can be found in the section on Methods.

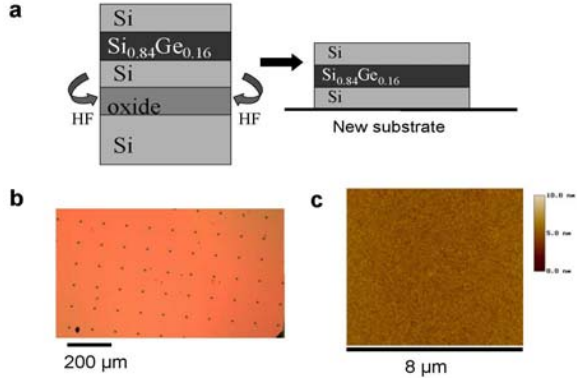

**Figure 2** Membrane formation. **a,** Schematic diagram of membrane formation and transfer. Etching of the oxide by HF releases the membrane, which is then transferred to a new substrate. **b,** Optical micrograph of a 1.8mm x 1mm part of a released 4mm x 4mm membrane of a strain-balanced 48 nm Si / 128 nm $Si_{0.84}Ge_{0.16}$ / 56 nm Si transferred in solution to an oxidized Si wafer. **c,** 8 μm by 8 μm AFM image of the same surface.

We transfer a 4 mm x 4 mm piece of the first sample onto a new silicon wafer (001). An optical micrograph of the released membrane is shown in Figure 2b. Membranes are very flexible while free, and can be adjusted on the new substrate until the water has evaporated. The transferred membrane shows few visible wrinkles or cracks over the entire 4 mm x 4 mm area of the film, suggesting the membranes are quite robust. Intermittent-contact mode atomic force microscopy (Fig. 2c) shows that the membranes are also flat on an extended scale, with no visible cross hatch. There was no change in the RMS roughness after release as measured by averaging over 5 squares of 2 μm x 2 μm on both the as-grown sample and following release. The RMS roughness of the as-grown film was 0.430nm and, after release, measured 0.425 nm. Roughness on this scale is typical of an unrelaxed SiGe film deposited on bulk Si.

To confirm that the release process causes elastic strain sharing in the released membrane, we made x-ray diffraction (XRD) reciprocal-space maps of the structure to determine the strain, thickness, and composition of the layer system both before and after its release. We measured the diffuse intensity around the (004) reflection to obtain a map containing information on strain and orientation in the as-grown state. Immediately following growth and before release, the system was coherent with the substrate (lattice matched, no dislocations, SiGe layer compressively strained), a conclusion that can be drawn from the alignment of the theta/2-theta peaks of the Si handle wafer and Si film as shown in Fig. 3a (an unprocessed SOI sample shows the same alignment). We confirmed structural coherency using maps of (224) reflections, which contain components of in- and out-of-plane lattice constants. Dislocation driven film relaxation in the Si/SiGe (001) system manifests itself in significant mosaic broadening and the absence of thickness fringes. The narrow Si and SiGe film diffraction peaks with strong thickness fringes in our samples support the lack of dislocation driven relaxation in the SiGe film. We fit theta/2-theta line scans to determine the composition of the SiGe layer and all layer thicknesses.

We were able to perform identical XRD measurements on the membrane after transfer, as shown in Figure 3b. Because the SiGe film is sandwiched between two Si layers of equal thicknesses, a flat membrane results, with the strain divided between the Si and SiGe. The new Si substrate is used to determine the zero-strain position of the Si peak in XRD in the same way as the handle wafer was used for the unreleased structure. The post-release XRD map of Figure 3b shows partially relaxed SiGe, and partially strained Si. The strain distribution matches the prediction of a simple elastic-strain sharing model discussed below. In the unreleased system, the Si handle-wafer peak and the Si film peak are aligned in the theta/2-theta scan. In the released and transferred membrane, the separation in theta between the Si substrate (new han-dle) and the Si membrane peak in the theta/2-theta scan is 0.091 ± 0.008°. The SiGe peak shifts by the same amount as the SiGe film relaxes. These peak separations correspond to a tensile strain in the Si layer in the released membrane of 0.30 ± 0.02 % and a compressive strain in the SiGe layer of the released membrane of 0.29 ± 0.02 %. In addition to demonstrating a strained Si layer, the XRD map shows omega peak widths of less than 0.03°. The widths of these diffraction peaks are an order of magnitude *narrower* than those typically observed in strained Si grown on a strain-graded SiGe substrate.[3] Line scans through the SiGe and Si films have thickness fringes from the finite film thicknesses before and after release as shown in Fig. 3a-b. The continued presence of thickness fringes and narrow peak widths are in agreement with dislocation-free relaxation.

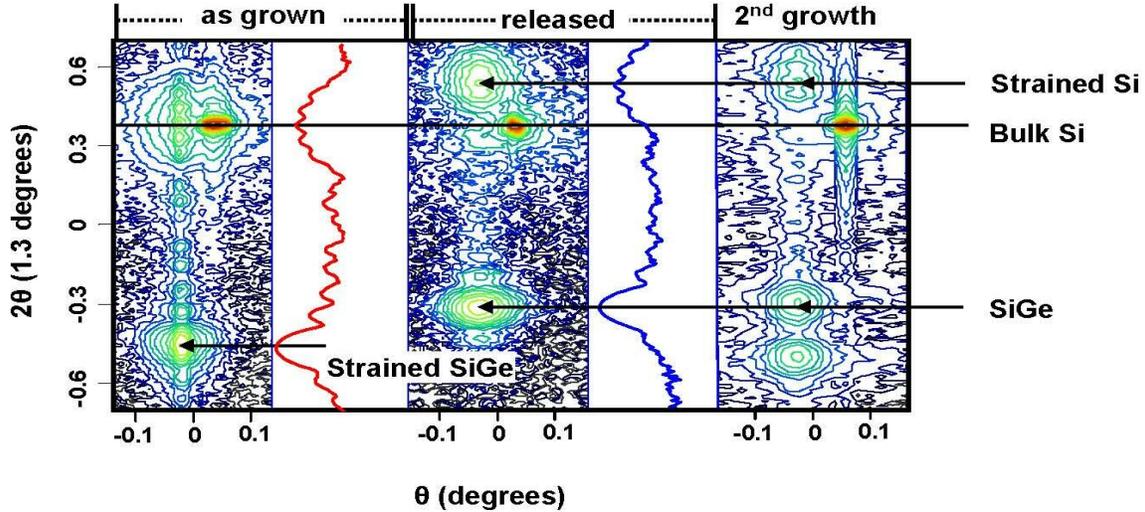

**Figure 3** XRD reciprocal-space maps of several Si/SiGe/Si membrane conditions. **a**, The Si/SiGe/Si film stack as deposited on an SOI substrate. The Si film peak and Si substrate peaks are perfectly aligned in 2 theta. A cut, as shown to the right of the map, through the Si/SiGe film peak was fit using the commercial package Bede Rads Mercury to determine layer thickness and Ge composition. **b,** The released membrane transferred to a new oxidized Si substrate. The Si film peak is shifted to a higher 2 theta angle from the Si substrate peak, indicating tensile strain in the Si membrane. The SiGe film peak has shifted to higher angle, consistent with relaxation of compressive strain. The line scan to the right of the map through the Si/SiGe/Si film peaks shows the continued presence of the thickness fringes following release. **c,** The transferred membrane after deposition of a second SiGe film with higher Ge concentration. The diffraction peak corresponding to this layer appears below the peak corresponding to the original SiGe film, indicating a higher compressive strain. The peaks for the original membrane do not move relative to the Si substrate, confirming no change in the strain state of the layers of the original membrane.

Figure 3c shows an XRD reciprocal-space map of the transferred membrane after depositing a second SiGe layer, which is chosen to have a higher Ge concentration. To grow this second layer, the usual wet chemical cleaning steps were followed, showing the robustness of the transferred membrane to processing. The peak from this second SiGe film is seen in theta just below the peak from the first SiGe film. The strained-Si film peak and the initial SiGe film peak have not moved in theta, indicating that the strain state of the pre-existing films *has not changed*. The lack of strain change is evidence that the bonding of the transferred membrane to the new substrate is good enough so that the membrane is not compliant. Of course, we can now again release the membrane system by etching the oxide, and the new membrane will again relax elastically, driven by the increased strain produced by the new SiGe layer.

The measured strain distribution in the Si/SiGe/Si membrane can be compared to the strain expected for an ideal elastic system with this thickness and initial strain state, using the relations

$$\varepsilon_{SiGe} = \varepsilon_m \frac{h_{Si} M_{Si}}{h_{SiGe} M_{SiGe} + h_{Si} M_{Si}} \quad (1)$$

$$\varepsilon_{Si} = -\varepsilon_m \frac{h_{SiGe} M_{SiGe}}{h_{Si} M_{Si} + h_{SiGe} M_{SiGe}} \quad (2)$$

where $\varepsilon_m$ is the mismatch strain, and $\varepsilon$, $M = \frac{E}{1-\nu}$, and h are the layer strain, biaxial moduli (in terms of Young's modulus and Poisson's ratio), and thicknesses of the Si and SiGe layers.[22] Equations (1) and (2) are derived from the force balance between the two layers and the requirement of lattice coherence across the interface.[22] This method of analysis is generalizable to many layer systems by adding a coherency equation corresponding to each new interface. For the layer system described here, with 104 nm of Si and 128 nm of $Si_{0.81}Ge_{0.19}$, the strain will have opposite sign for, and be distributed between, the two materials. The mismatch strain is $\varepsilon_m = 0.60\%$. The predicted peak separation corresponds to a tensile strain of $0.32 \pm 0.01$ % in the Si layers and a compressive strain of $0.27 \pm 0.01$ % in the SiGe layer. The predicted and measured strain values agree within the limits of uncertainty. The uncertainty in the predicted strain values comes from the uncertainties in measuring the thicknesses and composition of the layers.

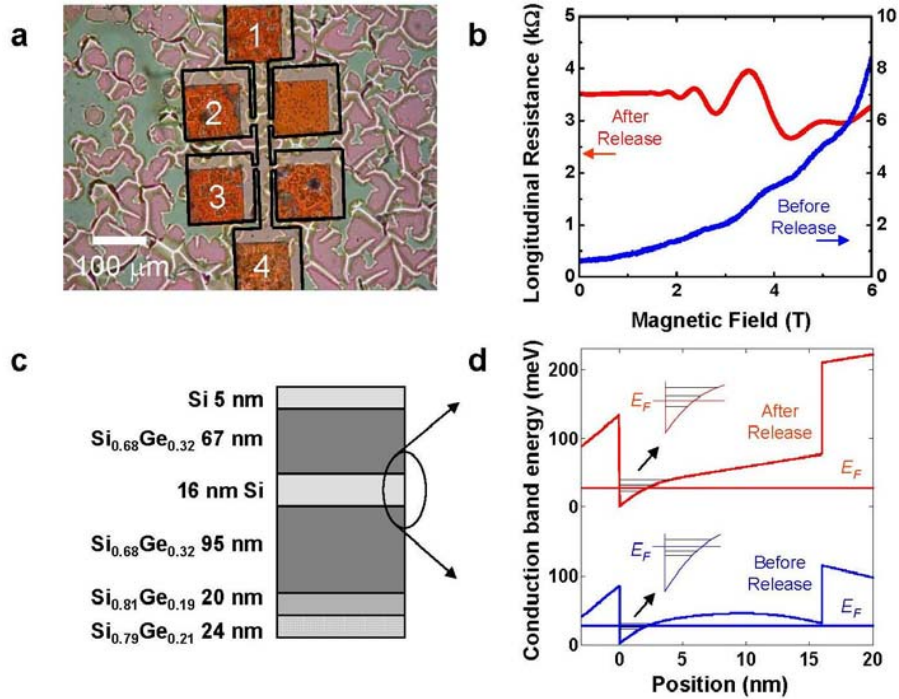

**Figure 4** Electron transport measurements in band-structure engineered elastically strained nanomembranes. **a,** Image of the released membrane and Hall bar with ohmic contacts; Hall bar is outlined for visibility. **b,** Longitudinal magnetoresistance before and after release. (Current from contacts 1 to 4. Voltage between contacts 2 and 3.) The strong positive magnetoresistance before release is indicative of many occupied subbands. The flat magnetoresistance and Shubnikov-de Haas oscillations after release are indicative of single subband occupation and two-dimensional transport. **c,** The measured heterostructure layers; the $Si_{0.81}Ge_{0.19}$ layer is phosphorous-doped at $2\times10^{18} cm^{-3}$; the upper $Si_{0.68}Ge_{0.32}$ layer is doped at $4\times10^{17} cm^{-3}$ for 34 nm of thickness, starting 11 nm above the quantum well. Before release, the silicon quantum well (circled) is tensile strained 0.76%. After release the strain increases to 0.99%. **d,** Calculations of the electronic states and the Fermi level in the before and after release quantum wells. Before release, two subbands are occupied, containing approximately 2/3 and 1/3 of the total charge respectively. After release, a single subband is occupied. Insets show subband details.

To confirm that the induced strain produces the desired electronic effects – in this case, the desired conduction band offsets – we grew a SiGe heterostructure containing a thin Si quantum well. Tensile-strained Si has a larger band offset with SiGe than unstrained Si does. Thus, better quantum well confinement and fewer occupied subbands after release would provide a demonstration of elastic strain sharing and band structure engineering. The initial substrate in this case was relaxed SiGe-on-insulator (SGOI), with a nominal composition of $Si_{0.80}Ge_{0.20}$. XRD measurements confirm that the actual composition is 21($\pm$2)% Ge and is 95(+5,-10)% relaxed, creating an initial tensile strain of 0.76% in the Si quantum well. The final structure, also from XRD, is 24 nm $Si_{0.79}Ge_{0.21}$, 20 nm $Si_{0.81}Ge_{0.19}$ P-doped at 2 x $10^{18}$ cm$^{-3}$, 95 nm undoped $Si_{0.68}Ge_{0.32}$, a 16 nm pure-Si quantum well, 11 nm undoped $Si_{0.68}Ge_{0.32}$, 34 nm $Si_{0.68}Ge_{0.32}$ P-doped at 4 x $10^{17}$ cm$^{-3}$, 22 nm undoped $Si_{0.68}Ge_{0.32}$, and a final cap layer of 5 nm Si (Fig. 4). We note here that the uncertainty in the starting SGOI substrate (composition and strain) is the largest uncertainty. Absolute determination of strain *and* composition using (224) maps suffers from angular uncertainty arising from the 0.2 degree mosaic spread of SGOI. Changes in strain alone (see below) are revealed much more precisely by XRD. Before release, Hall bars (300 μm x 20 μm) were etched into the heterostructure by a $SF_6$ reactive ion etch to a depth of 90 nm, and ohmic contacts were made using an evaporated Au/Sb alloy followed by a rapid thermal anneal. The patterned samples were then immersed in a buffered HF solution for 20 min, undercutting the entire structure, which was allowed to collapse in place onto the substrate. This in-place collapse results in some local buckling of the film as it elastically expands, as shown in Fig. 4a, but this effect contributes negligibly in x-ray measurements. X-ray diffraction shows an angular shift of the strained Si peak following release of 0.07° ± .005°, indicating an increase of strain in the Si quantum well from the initial 0.76% to 0.99% ± 0.01%, as expected for strain sharing.

Both before and after release, four-probe longitudinal magnetoresistance measurements were performed at a temperature of 2 K using ac lock-in techniques with 20 nA current bias at 11 Hz. As shown in Fig. 4b, the before-release longitudinal resistance shows a pronounced upward curvature typical of a quantum well with more than one occupied subband.[23] Weak Shubnikov-de Haas oscillations are superimposed in the before-release data and arise from the lowest occupied subband.[23] Following release, the elastic strain relaxation in the thick $Si_{0.68}Ge_{0.32}$ layers causes an increase in the tensile strain of the thin silicon quantum well, resulting in an increase in quantum well depth of approximately 50 meV. The longitudinal resistance after release shows a *flat* magnetic-field dependence for small magnetic fields and progressively larger Shubnikov-de Haas oscillations at higher fields. Such a curve, with low or zero curvature at zero field, is a hallmark of two-dimensional electronic transport in a single subband[24] and demonstrates that the elastic relaxation does indeed create the desired electronic-band-structure effects. Fig. 4d (top) shows the calculated Fermi level and subband energies after release, using the 2DEG electron density, the observation of single-subband occupation from Fig. 4b, and band offsets from reference 25 as input parameters. Fig. 4d (bottom) shows the subband energies and Fermi level before release, with the multiple-subband occupation resulting solely from the smaller tensile strain in the silicon layer before the elastic strain relaxation of the entire heterostructure. It is useful to note that the electronic transport is not sensitive to the bowing observed in the membrane, as shown in Fig. 4a, and, of course, that the membrane is indeed flexible. The insensitivity of transport to this bowing is in part due to the four-probe measurement geometry, so that any distortions at the contacts are unimportant, and it is due in part to the small area fraction of the bowed regions. As discussed, the curvature in this sample results from dropping the membrane in place, in order to facilitate our before-and-after comparison. Transfer of a membrane from one substrate to another would result in no buckles (as in the first example in this paper).

For the purposes of this paper, we used thick Si layers and low SiGe composition to transfer relatively small strains to Si as a proof of principle. XRD confirms that release causes *elastic* strain relaxation, and comparison with simple theory demonstrates ideal strain sharing behavior. The strain transferred to the Si can be greatly increased by decreasing the total Si thickness and/or increasing the Ge composition of the alloy. We have grown Si/SiGe/Si structures with such higher Ge compositions in the alloy layer and with thinner relative Si thickness, to transfer more strain to the Si. As bonding is not essential to maintain strain in the system, a very wide range of possibilities exists for the choice of the new substrate; good adhesion be-tween the layers in the structure itself is of course necessary. We expect that membrane transfer to flexible substrates is possible, and we also believe the concept presented here can be applied directly to other materials systems for which a release layer can be incorporated into the structure.

This research was supported by DOE, NSF-MRSEC, AFOSR, NSF-ITR, and ARDA.

**Methods**

Membranes are released using wet chemical etching with hydrofluoric acid (HF). To increase the rate of membrane release, we generally pattern an array of holes using conventional photolithography and etch down to the oxide using reactive ion etching to provide access for the HF. Before release, the membrane-on-insulator is cleaned to remove any remaining photoresist, and the oxide is etched in a 49% HF solution for 10 – 90 minutes, depending on the separation of the access holes. We choose the separation of the holes on the basis of the desired HF etch time. A film system with 5 μm holes and a 200 μm pitch requires a 90 min etch in 49% HF. We have made membranes in sizes from ½ x ½ cm down to 5 x 5 μm; both larger ones and smaller ones are possible. No holes are used for small-size membranes.

After the buried oxide is completely removed, the membrane is loosely bound to the surface of the Si handle wafer. The piece is transferred from the HF to a DI water beaker. The membrane can be completely released from the starting substrate with agitation or the addition of a solvent at which point it will float on the surface of the water, where it is easily picked up by the new desired substrate. Figure 2a shows a schematic diagram of the film structure and release process. So far we have transferred membranes to several new substrates, as mentioned above. Membranes can, of course, also be transferred to curved substrates, either rigid or flexible.

For the purposes of this paper, we describe membranes transferred to new bulk-Si substrates terminated with 100 nm of thermal $SiO_2$. After drying, the membrane sticks weakly to the new substrate. To obtain a strong bond between the membrane and the substrate, we anneal the membrane without pressure on a hot plate for 5 min at 100 °C followed by 5 min at 500 °C. After this heat treatment, the membrane remains flat and can survive chemical processing (piranha, AHP, HF) with no damage. These bonded membranes can be used as substrates for further growth, as we have demonstrated by growing a second SiGe film of higher Ge composition via CVD on a released, bonded membrane that was annealed and chemically cleaned. Hence a multiple-growth-and-release process is feasible.

**References:**

1. Ieong, M., Doris, B., Kedzierski, J., Rim, K. & Yang, M. Silicon device scaling to the sub-10-nm regime. *Science* **306**, 2057-2060 (2004).
2. Rim, K., Hoyt, J. L. & Gibbons, J. F. Fabrication and analysis of deep submicron strained-Si N-MOSFET's. *IEEE Trans. Electron Devices* **47**, 1406-1415 (2000).
3. Mooney, P. M. & Chu, J. O. Heteroepitaxy and high-speed microelectronics. *Annu. Rev. Mater. Sci.* **30**, 335-362 (2000).
4. Fitzgerald, E. A. *et al.* Relaxed $Ge_xSi_{1-x}$ structures for III-V integration with Si and high mobility two-dimensional electron gases in Si. *J. Vac. Sci. Technol. B* **10**, 1807-1819 (1992).
5. Ismail, K. *et al.* Identification of a mobility-limiting scattering mechanism iin modulation-doped Si/SiGe heterostructures. *Phys. Rev. Lett.* **73**, 3447-3450 (1994).
6. Monroe, D., Xie, Y. H., Fitzgerald, E. A., Silverman, P. J. & Watson, G. P. Comparison of mobility-limiting mechanisms in high-mobility $Si_{1-x}Ge_x$ heterostructures. *J. Vac. Sci. Technol. B* **11**, 1731-1737 (1993).
7. Lo, Y. H. New approach to grow pseudomorphic structures over the critical thickness. *Appl. Phys. Lett.* **59**, 2311-2313 (1991).
8. Brown, A. S. Compliant substrate technology: Status and prospects. *J. Vac. Sci. Technol. B* **16**, 2308-2312 (1998).
9. Hobart, K. D. *et al.* Compliant substrates: A comparative study of the relaxation mechanisms of strained films bonded to high and low viscosity oxides. *J. Electron. Mater.* **29**, 897-900 (2000).
10. Yin, H. *et al.* Buckling suppression of SiGe islands on compliant substrates. *Journal of Applied Physics* **94**, 6875-6882 (2003).
11. Ejeckam, F. E., Lo, Y. H., Subramanian, S., Hou, H. Q. & Hammons, B. E. Lattice engineered compliant substrate for defect-free heteroepitaxial growth. *Appl. Phys. Lett.* **70**, 1685-1687 (1997).
12. Mooney, P. M., Cohen, G. M., Chu, J. O. & Murray, C. E. Elastic strain relaxation in free-standing SiGe/Si structures. *Appl. Phys. Lett.* **84**, 1093-1095 (2004).

13. Jones, A. M. *et al.* Long-wavelength InGaAs quantum wells grown without strain-induced warping on InGaAs compliant membranes above a GaAs substrate. *Appl. Phys. Lett.* **74**, 1000-1002 (1999).
14. Cohen, G. M., Mooney, P. M., Paruchuri, V. K. & Hovel, H. J. Dislocation-free strained silicon-on-silicon by in-place bonding. *Appl. Phys. Lett.* **86**, 251902 (2005).
15. Damlencourt, J.-F. *et al.* Paramorphic Growth: A new approach in mismatched heteroepitaxy to prepare fuly relaxed materials. *Jpn. J. Appl. Phys.* **38**, L996-L999 (1999).
16. Boudaa, M. *et al.* Growth and characterization of totally relaxed InGaAs thick layers on strain-relaxed paramorphic InP substrates. *J. Electron. Mater.* **33**, 833-839 (2004).
17. Demeester, P., Pollentier, I., De Dobbelaere, P., Brys, C. & Van Daele, P. Epitaxial lift-off and its applications. *Semicond. Sci. Technol.* **8**, 1124-1135 (1993).
18. Menard, E., Lee, K. J., Khang, D.-Y., Nuzzo, R. G. & Rogers, J. A. A printable form of silicon for high performance thin film transistors plastic substrates. *Appl. Phys. Lett.* **84**, 5398-5400 (2004).
19. Yablonovitch, E., Hwang, D. M., Gmitter, T. J., Forez, L. T. & Harbison, J. P. Van der Waals bonding of GaAs epitaxial liftoff films onto arbitrary substrates. *Appl. Phys. Lett.* **56**, 2419-2421 (1990).
20. Langdo, T. A. *et al.* SiGe-free strained Si on insulator by wafer bonding and layer transfer. *Appl. Phys. Lett.* **82**, 4256-4258 (2003).
21. Moriceau, H. *et al.* New layer transfers obtained by the SmartCut Process. *J. Electron. Mater.* **32**, 829-835 (2003).
22. Freund, L. B. & Suresh, S. *Thin film materials* (Cambridge University Press, Cambridge, 2003).
23. van Houten, H., Williamson, J. G., Broekaart, M. E. I., Foxon, C. T. & Harris, J. J. Magnetoresistance in a GaAs/$Al_xGa_{1-x}As$ heterostructure with double subband occupancy. *Phys. Rev. B.* **37**, 2756-2758 (1988).
24. Beenakker, C. W. J. & van Houten, H. Quantum transport in semiconductor nanosturctures. *Solid State Physics* **44**, 1 (1991).
25. Schäffler, F. High-mobility Si and Ge structures. *Semicond. Sci. Technol.* **12**, 1515-1549 (1997).